# Continuum-atomistic simulation of picosecond laser heating of copper with electron heat capacity from *ab initio* calculation

Pengfei Ji and Yuwen Zhang[1]

Department of Mechanical and Aerospace Engineering

University of Missouri

Columbia, MO 65211, USA

## Abstract

On the basis of *ab initio* quantum mechanics (QM) calculation, the obtained electron heat capacity is implemented into energy equation of electron subsystem in two temperature model (TTM). Upon laser irradiation on the copper film, energy transfer from the electron subsystem to the lattice subsystem is modeled by including the electron-phonon coupling factor in molecular dynamics (MD) and TTM coupled simulation. The results show temperature and thermal melting difference between the QM-MD-TTM integrated simulation and pure MD-TTM coupled simulation. The successful construction of the QM-MD-TTM integrated simulation provide a general way that is accessible to other metals in laser heating.

## 1. Introduction

In the past decades, ultrashort laser material processing has been an increasingly hot topic and has drawn lots of attentions from researchers. Numerical modeling and simulation of the laser material interaction ranging from two temperature model (TTM) which assumes the laser energy is firstly absorbed by electron subsystem[1–4], molecular dynamics simulation (MD) which describes the laser heating by scaling the atomic velocity[5], to the quantum mechanics (QM) approach which introduces the Fermi-Dirac distribution of the instantly increased the electron temperature and performs the electron phonon interaction subsequently[6,7]. Furthermore, there are several MD-TTM combined schemes, which describes the atomic motion by using MD and models the electron subsystem by using a continuum energy equation[8–12]. However, the thermophysical parameters of electron required in the energy equation, such as the electron heat capacity $C_e$, is necessary in these combined MD-TTM schemes. Due to the strong thermal excitation of the electron subsystem, $C_e$ varies greatly with electron temperature $T_e$.

Precise knowledge of the electron heat capacity helps to better explain the measurement of material temperature upon laser irradiation, facilitates the theoretical and numerical investigations of melting, vaporization and sublimation. The *ab initio* calculation provides a novel way of obtaining the electron temperature-dependent $C_e(T_e)$[13–16]. There are limited number of work performing the *ab initio* calculation of $C_e(T_e)$ and plugins it directly into the MD-TTM simulation. Lin and Zhigilei carried out *ab initio* calculations to study $C_e$ for a serials of metals[16], but the variations of electron density of states (EDOS) were not taken into account. Whereas, the EDOS differs a lot at high electron temperature[14]. Bevillon *et al.*[13] calculated the electron behavior of metals under electron-phonon nonequilibrium resulting from laser irradiation, and the free-electron properties were determined at atomic level. This letter reports the first effort of $T_e$ dependent $C_e$ for copper through the computation in the following section.

---

[1] Corresponding author. Email: zhangyu@missouri.edu

## 2. Computational details

The present letter concentrates on the determination of $C_e(T_e)$ of copper via QM and its implementation into the upper scale of MD-TTM coupled simulation to study the ultrashort laser heating of copper film. The electron heat capacity, $C_e$, of copper is computed by taking derivative of the internal energy of the electron subsystem $E_e$ with respect to $T_e$, namely $\partial E_e/\partial T_e$. Since the EDOS $g$ is $T_e$-dependent, the electron heat capacity[16] can be rewritten as

$$C_e(T_e) = \int_{-\infty}^{\infty} \left(\frac{\partial f}{\partial T_e}g + f\frac{\partial g}{\partial T_e}\right)\varepsilon d\varepsilon \qquad (1)$$

where $f$ is the Fermi-Dirac distribution function $1/(e^{\frac{\varepsilon-\mu}{k_B T_e}}+1)$, which is a function of $T_e$ and energy level $\varepsilon$. $\mu$ is the chemical potential at given $T_e$. In this letter, the plane wave density functional theory (DFT) code ABINIT[17] was used to perform the massive parallelism calculations of the electron temperature-dependent $g$ and $f$[18]. By substituting $g$ and $f$ and their derivatives with respect to $T_e$ into Eq. (1), $C_e(T_e)$ was determined; this approach has not been reported before. The nucleus and core electrons of copper were modeled by the projector-augmented wave (PAW) atomic data[19], which took 11 valence electrons per atom. The local density approximation (LDA) functional developed by Perdew and Wang[20] was included for the exchange and correlation functional. The Brillouin zone was sampled by using the Monkhorts-Pack method[21]. Convergence test results showed the $18 \times 18 \times 18$ $k$-point grids and cutoff energy of 32 Ha are sufficient to obtain converged energy. Face centered cubic crystal of copper was established. The test result of the PAW atomic data showed the lattice constant of $3.662$ Å with a relative error of 1.8 % to the experimental value 3.597 Å at 300 $K$[22], which demonstrated the reliability of the PAW atomic data. In the next step, 50 bands per atom was set to ensure the maximum occupation of electrons. The lattice temperature $T_l$ was kept at room temperature (300 $K$), while $T_e$ was set at different values range from 300 to 50,000 $K$.

The energy equation for the electron subsystem is mathematically expressed as:

$$C_e(T_e)\frac{\partial T_e}{\partial t} = \nabla(K_e\nabla T) - G(T_e - T_l) + S(x,t) \qquad (2)$$

where $K_e$ and $G$ represent the electron thermal conductivity and electron-phonon coupling factor. Both $K_e$ and $G$ are treated as constant in the present work, which are $400\ W/(mK)$ and $1.0 \times 10^{17}\ W/(m^3 K)$[24], respectively. Similar treatment of $K_e$ and $G$ are seen in Ref.[9] for aluminum. $t$ represents time, and $x$ denotes the direction of laser incidence, which is perpendicular to $y$-$z$ plane. $S$ is the source term of incident laser, whose density $S(x,t)$ is given as a temporal- and spatial-dependent (simplified as one dimensional) Gaussian profile

$$S(x,t) = J(1-R)(t_p L_{op}\sqrt{2\pi})^{-1} e^{-x/L_{op}} e^{-(t-t_0)^2/2t_p^2} \qquad (3)$$

where $J = 21,333 \, J/m^2$ is the laser fluence. $R = 0.4$ is the reflectivity and $L_{op} = 14.29 \, nm$ is the optical penetration depth, which is chosen for an incident laser with laser wavelength of $\sim 320 \, nm^{25}$. $t_0 = 50 \, ps$ is the temporal center point of the laser beam. $t_p = 10 \, ps$ is the full width of laser pulse at half maximum intensity.

For the lattice subsystem, the motion of atoms is determined by the

$$m_i \, d^2\boldsymbol{r}/dx^2 = -\nabla U + \xi m_i \boldsymbol{v}_i^T \tag{4}$$

where the first term in the right side is the spatial derivative of interatomic potential $U$. The embedded atom method (EAM) potential of copper[26] is adopted in the present work. $m_i$ is the mass of an atom. $\boldsymbol{r}$ is the atom position at given time. $\xi$ in the last term in right side is defined as $\frac{1}{n_t}\sum_{k=1}^{n} GV_N(T_e^k - T_l) / \sum_{j=1}^{N_v} m_j(\boldsymbol{v}_j^T)^2$, which is originally proposed by Inanov and Zhigilei[8] to couple the thermal energy transferring from the electron subsystem to the lattice subsystem. $\boldsymbol{v}_i^T$ is the thermal velocity of atom $i$. $T_e^k$ is the average electron temperature in each MD time step. The continuum region is divided into $N = 888$ cells with $n$ (variable) atoms in each cell to solve Eq. (2) by using explicit finite difference method (FDM). In order to satisfy the von Neumann stability criterion, the MD time step is set as several times of the FDM time step, namely, $\Delta t_{FDM} = \Delta t_{MD}/n_t < 0.5\Delta x_{FDM}^2 C_e/K_e^{27}$. In the current letter, a conservative estimation was made by choosing $\Delta t_{MD}$ as $1 \, fs$ and $\Delta t_{FDM}$ as $0.005 \, fs$.

By combing Eqs. (2)-(5) to solve the laser energy deposition in the electron subsystem and atomic motion in the lattice subsystem, a framework QM-MD-TTM integrated simulation is constructed. The QM-MD-TTM integrated simulation was performed from the revision of the TTM part in the IMD package[28,29].

The entire simulation was performed in three sequential stages. Initially, the entire system was equilibrated at room temperature ($300 \, K$) in terms of canonical ensemble (the 1$^{st}$ stage) to keep the lattice temperature constantly for $5 \, ps$. Subsequently, microcanonical ensemble (the 2$^{nd}$ stage) was started to verify whether the lattice subsystem was well equilibrated for another $5 \, ps$. Meanwhile, the electron temperature was prepared at $300 \, K$ since the beginning of the simulation. The QM-MD-TTM integrated simulation (the 3$^{rd}$ stage) started at $10 \, ps$ and lasted for $240 \, ps$. The entire system was established with $578.3840 \, nm$, $3.6149 \, nm$ and $3.6149 \, nm$ in $x$-, $y$- and $z$- directions, respectively. It should be noted that the thickness of copper film is $347.0304 \, nm$. There were two vacuum spaces with thickness of $173.5152 \, nm$ above (occupying 30% of the entire length in $x$- direction) and $37.8384 \, nm$ below the film (occupying 10% of the entire length in $x$- direction), which were set to allow the film to expand during and after laser irradiation. The total number of copper atoms contained in the system was 384,000. Meanwhile, the treatment of $C_e$ in terms of experimental result $C_e = \gamma T_e$ (where $\gamma$ is $96.8 \, J/(m^3 K^2)^{23}$) in Eq. (2) are performed in the MD-TTM simulation to compare the effects of *ab initio* result.

## 3. Results and discussion

### 3.1 The $T_e$ dependent $C_e$

The calculated $g$ at $T_e$ of $300 \, K$, $10,000 \, K$, $30,000 \, K$ and $50,000 \, K$ are shown in Fig. 1(a). In order to take the comparisons of Fermi-Dirac distribution $f$ (see Fig.1(b)) at corresponding temperatures into account, the zero point of the horizontal axis was set as the Fermi energy $\varepsilon_F$. As seen in Eq. (1), the $T_e$-dependent $C_e$ is only determined by the part that $T_e$ derivate of the $g$ and $f$ are not equal to zero. With the increasing $T_e$, $g$ moves toward the lower energy region. In order to conserve the number of valence electrons, the chemical potential $\mu$ moves to higher value in response to the overall shift of $g$ towards lower energy region, which is reflected as $f$ moves to the higher energy region. Table 1 lists the relative change of $\mu - \varepsilon_F$ to $T_e$. It can be seen that the $\mu$ increases at greater $T_e$, which leads to the translation of the $f$ to higher energy region. The translation is reflected in Fig. 1(b). Moreover, the increase of $\mu$ at greater $T_e$ brings effect of $C_e(T_e)$.

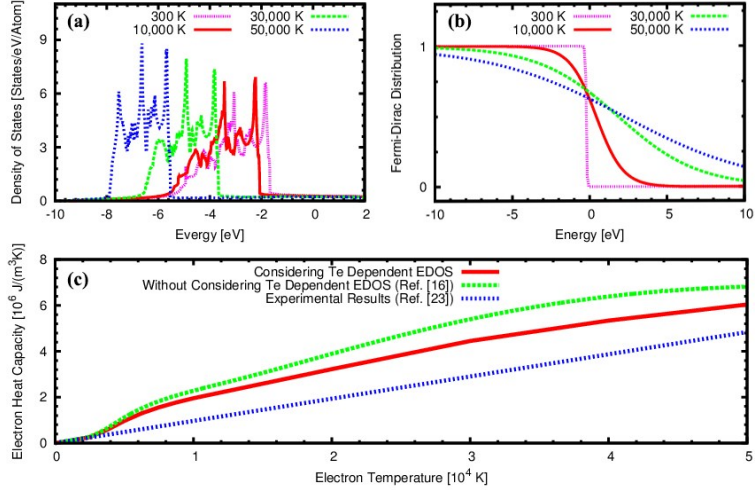

Fig. 1 (a) Electron density of states and (b) Fermi-Dirac distribution for electrons at $300\ K$, $10,000\ K$, $30,000\ K$ and $50,000\ K$. (c) $T_e$-dependent $C_e$ from *ab initio* calculation with and without considering the $T_e$-dependent EDOS [16] and experimental result [23].

**Table 1** Variation of $\boldsymbol{\mu - \varepsilon_F}$ at different $\boldsymbol{T_e}$

| $T_e$ (K) | $\mu - \varepsilon_F$ (eV) |
|---|---|
| 10,000 | 0.4266 |
| 20,000 | 1.2756 |
| 30,000 | 1.8427 |
| 40,000 | 2.1332 |
| 50,000 | 2.2180 |

The calculated $C_e(T_e)$ is shown in Fig. 1 (c). Meanwhile, *ab initio* calculation without considering the $T_e$-dependent $g$ [16], and experimental result[23] $C_e = \gamma T_e$ (where $\gamma$ is 96.8 J/$(m^3 K^2)$[23]) are also drawn in Fig. 1(c) for comparison. Comparing $C_e(T_e)$ obtained in this letter with $C_e(T_e)$ from Refs. [16] and[23], the three agree well when $T_e$ is below $1,500\ K$. However, with the continuous increase of $T_e$, the two *ab initio* calculated $C_e$ increase faster than the experimental result. As aforementioned in Fig. 1(a), because $g$ towards lower energy region with the increase of $T_e$, the calculated $C_e$ is lower than that without considering variation of $g$ at given $T_e$. Therefore, the overestimation of $C_e$ will lead to lower $T_e$ response in MD-TTM coupled simulation. Similarly, because of $C_e$ at high $T_e$ are greater than experimental result, after enormous amount of ultrafast laser energy deposition, the $T_e$ response in the MD-TTM simulation by using the *ab initio* calculated $C_e(T_e)$ will be lower than those using experimental $C_e$, which will be seen in the subsequent simulation results.

### 3.2 Spatial and temporal distribution of $T_l$, $T_e$ and $\rho$

The temporal evolutions of lattice temperature $T_l$ and electron temperature $T_e$ distributions normalized along the laser incident direction is shown in Fig. 2. The left side of the horizontal $x$-axis represents front end of the film surface, while right side represents rear end of the cooper film surface. As seen in the insets of Figs. 2(a) and (b), $T_l$ appear horizontally at the $10\ ps$, which indicate the lattice subsystems have been well equilibrated at room temperature. For the reason that the electron-electron interaction is at the timescale of femtosecond and the electron-phonon interaction is at the timescale of picosecond scale, even though the laser intensity has reached the maximum value, there is great temperature difference between $T_e$ and $T_l$. At $50\ ps$, $T_e$ in Figs. 2(a) and 2(b) show the greatest value among all the sampled electron temperature profiles. When $T_e$ is greater than $2,700\ K$, because of $C_e$ calculated in this letter is much greater than that from experiment at given $T_e$, the maximum $T_e$ ($38,000\ K$) shown in Fig. 2(a) is lower than that ($44,200\ K$) shown in Fig. 2(b) after the same amount of laser energy is deposited into the electron subsystem. When it comes to $70-85\ ps$, $T_l$ at the region (which locates $x < 0.4$ in the inset figures of Fig. 2) present the highest values of all the six lattice temperature distributions. $T_e$ at the front side of the film decreases after $50\ ps$, which results from the thermal energy transports from electron subsystem to the lattice subsystem is greater than the deposited laser energy per unit time. Moreover, it can be observed that $T_e$ at the middle region of the film ($0.35 < x < 0.65$) shows the highest temperature at $70\ ps$ than those of subsequent time points (at $85\ ps$ and $250\ ps$), which is induced by the faster electron heat conduction than the electron-phonon coupled thermal energy transfer. The left shift of the $T_e$ and $T_l$ peaks at the front surface reflects the thermal expansion of the film.

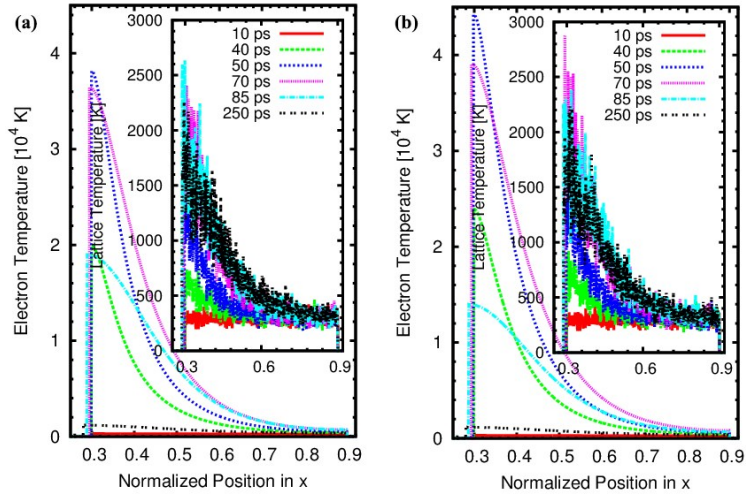

Fig. 2 Spatial distribution of $T_e$ and $T_l$ from (a) the calculation of QM-MD-TTM integrated simulation and (b) MD-TTM coupled simulation by implementing the experimental $C_e$.

In order to get a further sight into the phase change of behind the $T_e$ and $T_l$ evolutions, spatial and temporal distribution of the density $\rho$ of the copper film are illustrated in Fig. 3. It can be seen that the density of copper before laser irradiation stabilizes at 8.94 $g/cm^3$ (slightly below density of solid copper 8.985 $g/cm^3$ at 293 $K^{30}$). Viewing from the time and position diagram since the point of laser irradiation, the density in front of the front surface of the film develops to lower values than those in the middle and rear end of the film. After 50 $ps$, density distribution at the front of the film gradually expands towards the vacuum space. As reported in[30], copper density drops from 8.35 $g/cm^3$ at the melting point to lower values due to expansion. There are some purple spots appearing at both front and rear of the copper film, which indicate the small fractions of copper atoms spread into air space caused by laser irradiation. A gradually developing region with density clearly lower than those at middle and rear regions of the film appears in Fig. 3, which forms a melting boundary between liquid and solid copper. Around 75 $ps$, it can be seen the starting point of the melting boundary is slightly earlier for the case in Fig. 2(b) than that in Fig. 2(a), which can be sourced from the higher $T_e$ induced by lower experimental $C_e$ after the deposition of the same amount of laser energy.

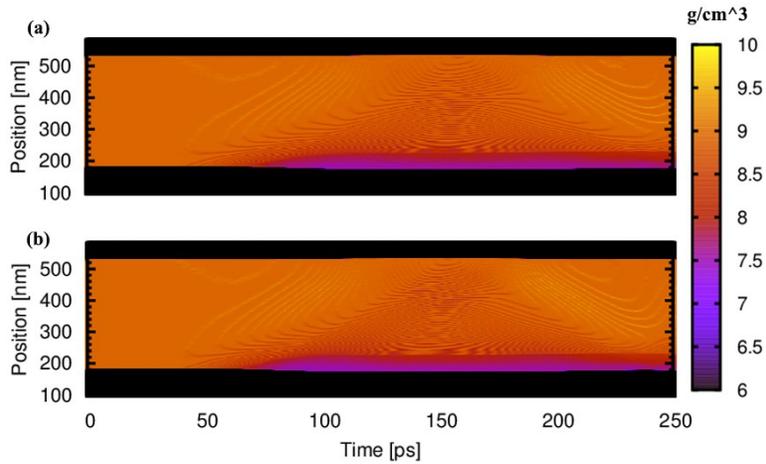

fig. 3 Comparison of the temporal and spatial distribution of density $\rho$ from (a) the calculation of QM-MD-TTM integrated simulation and (b) MD-TTM coupled simulation by implementing the experimental $C_e$.

In addition, there are also some wave patterns and lines traveling back and forth with the advance of time. From the pattern distributions, it can be seen some of the waves show higher densities than the density of copper film before laser irradiation. While some of the waves render obviously lower densities because of laser induced thermal expansion. By taking the one of relatively straight lines (in bright yellow) from 75 $ps$ to 100 $ps$ for an example, the traveling speed of compressed point is approximately at 5 $km/s$, which is comparable to the longitudinal speed of sound for copper 5.01 $km/s^{31}$. Whereas, for the wave patterns representing smaller density than that at room temperature, the traveling speed of expanded point is much slower (please see the placid dark waves). When the bright wave travels to the rear side of the copper film, it disappears in terms of reflected waves, as a result of slightly expansion of the rear side. Meanwhile, there are also expanded points generated under the melting boundary traveling to the rear side of the film. The two waves coming from the front and rear sides meet at the position of ~420 $nm$ at

~150 $ps$. Observing the number of waves that the moment of the two kinds of waves colloid, there are denser waves in Fig. 3(b) than that in Fig. 3(a) at 150 $ps$. The two waves travel back towards their incoming direction after collision of the two waves. The wave traveling back to the front side of the film produces density dilution in the melted region by penetrating the melting boundary (at $180 - 200\ ps$). A new group of points compose bright waves, as a result of the collided waves traveling back to the rear side.

After 125 $ps$, it can be seen that the clear and stable boundary between the solid copper and melted copper. Therefore, it can be concluded that a melting boundary has reached the steady state. Based on the thickness of the melted region in Fig. 3, for both cases in Figs. 2(a) and 2(b), the melting depth of the given laser irradiation of copper film are determined around 50 $nm$. Even though $C_e$ are not equal at given $T_e$ (as seen in Fig. 1(c)), for the reason that the same amounts of laser energy are deposited into the laser film, the final melting depths are no significant difference after sufficient long time of electron-phonon relaxation.

## 4. Conclusions

The ultrafast laser interaction with copper film is simulated using the QM-MD-TTM integrated framework. Owing to the accuracy of QM approach from pure electronic structure calculation and no need of empirical precondition, *ab initio* calculated $C_e$ is implemented into MD-TTM couple simulation. It can be concluded that even though significant $T_e$ difference between the QM-MD-TTM simulation and conventional MD-TTM coupled simulation by substituting $C_e$ from experimental result[16], there is slight difference between the final melting depths of the laser irradiated films. Nevertheless, the slower starting point of melting and smaller number of density waves reveal the importance of precisely determine $C_e$. The current letter empowers the inclusion of the *ab initio* determination for $C_e$ in the QM-MD-TTM integrated simulation and paves a new way to model the multiscale simulation of laser material processing. Besides the investigation of ultrafast laser heating of copper film, the successful construction of the QM-MD-TTM integrated simulation provide a general way that is accessible to other metals.

## Acknowledgments

Support for this work by the U.S. National Science Foundation under grant number CBET-133611 is gratefully acknowledged.